\documentstyle[psfig]{article}
\begin{document}
\newcommand{\ket}[1]{\left|#1\right\rangle}
\newcommand{\LK}{\ket{LK}}
\newcommand{\lke}{\ket{LKE}}
\newcommand{\uk}{\ket{\uparrow}}
\newcommand{\dk}{\ket{\downarrow}}
\newcommand{\rk}{\ket{\rightarrow}}
\newcommand{\lk}{\ket{\leftarrow}}

\title{Localisation and Nonlocality in Compound Energy-Momentum Eigenstates}
\author{Michael Simpson \\
         \\
         {\it Department of Mathematics} \\
         {\it University of Western Australia} \\
         {\it Nedlands, Western Australia 6907} \\
              email: simpson@maths.uwa.edu.au }
\date{November 1995}

\maketitle

\setlength{\parindent}{0em}

	A thought experiment considering conservation of energy and momentum
for a pair of free bodies together with their internal energy is used to
show the existence of states that have localised position while being
eigenstates of energy and momentum.  These states are applicable to all
varieties of physical bodies, including planets and stars in free motion
in the universe.  The states are compound entanglements
of multiple free bodies in which the momenta of the bodies are
anticorrelated so that they always sum to zero, while their total
kinetic energy is anticorrelated with their internal energies, so
the total is a constant, $E$. The bodies are relatively localised while
the total state has well-defined energy and momentum.  These states
do not violate Heisenberg uncertainty because the total centre of mass
is not localised, hence the states naturally describe whole universes
rather than isolated systems within a universe.  A further property
of these states, resulting from the form of the entanglement, is that
they display nonlocality in the full sense of signal transmission rather
than the more restricted Bell sense.


\setlength{\parindent}{5em}
\setcounter{section}{-1}

\section{Introduction}

	One of the fundamental results of quantum theory is the
Heisenberg uncertainty principle.  If two observables are represented
by non-commuting operators then it is not possible for those observables
to simultaneously have precise values.  The first example normally
mentioned is position-momentum uncertainty:  the more precisely position
is known, the less precisely momentum will be known, and vice-versa.
This leads to the expectation that a body in free space cannot
simultaneously occupy a momentum eigenstate and be localised, that is,
have its position defined to within narrow limits.

	The purpose of the present work is to show that this expectation,
while strictly correct, is misleading in an important way.  It is
possible for bodies to be localised and yet at the same time to occupy
a momentum eigenstate.  The smallest state permitting this requires two
bodies whose momenta are anticorrelated, that is, are precisely equal
and opposite.  In this case the two bodies are relatively localised even
though jointly occupying an eigenstate of their total momentum having
eigenvalue $0$.  The centre of mass state is completely unlocalised,
as demanded by the uncertainty relations.  This state is essentially
the one described by Einstein, Podolsky and Rosen in their
famous paper.\cite{epr}

	Because the centre of mass is not localised, the above state
would appear to be of no practical importance beyond specialised uses:
the bodies are localised relative to each other, but are not localised
relative to the rest of the universe.  The present paper shows, however,
that a state can be constructed containing any arbitrary number of
bodies, with all bodies localised relative to each other, and with
total momentum $0$.  If such a state described the entire universe,
the lack of absolute centre of mass localisation would be irrelevant
as there would be no external observer, and observers within the universe
would find all observable bodies localised.

	The paper also shows that the two-body state can be extended in a
different way, so that the total kinetic energy of the bodies is placed in
correlation with the internal energy of the bodies so that their sum is a
constant value $E$, thus creating an eigenstate of total energy.  This can
be done while retaining the relative localisation and precisely defined
total momentum, opening the possibility that the universe as a whole
occupies an energy-momentum eigenstate.  It is not proved, however, that
this can be extended to arbitrarily large numbers of bodies, although it
appears highly unlikely that this extension would fail.

	Localised energy-momentum eigenstates have a further pair of
remarkable properties.  The first is that they violate the well-known
dictum that interference cannot continue if we are in a position to
determine which of the interfering states the system occupies.
Localisation is a phenomenon resulting from interference between
free-body momentum states, but these states are now in correlation
with the internal energy states of the bodies.  If the internal energy
were measured the momentum state of the system would be known, yet
the interference remains.

	The second property derives from the first.  If we actually
measure the internal energy of the bodies each of them will be placed
in a momentum eigenstate, a state that is completely unlocalised.
This loss of localisation would instantly affect the whole universe,
and would be readily observable, allowing it to be used as a means of
superluminal signal transmission.  These localised energy-momentum
eigenstates are therefore also {\it nonlocal}, in the entirely different
sense of permitting physical influences to travel across space.
Moreover, unlike the Bell nonlocality,\cite{desp} these influences
appear to be usable to transmit signals, and so may profoundly alter
our present ideas about the relationship between quantum theory and
special relativity.

\section{A Thought Experiment}

	Consider a solid body.  Make the normal separation between the
centre of mass state and the internal state.  Initially, the internal
state is an excited energy eigenstate of energy $E$ at least sufficient
to eject one electron from the body.  Such states are also momentum
eigenstates of eigenvalue $0$.  At a later time one electron has been
emitted from this body.  (It might be useful to picture the body as an
electron gun before and after emitting.)  The state is now composed of
three parts:  the internal state of the body, lacking one electron; the
state describing the relative position of the body and the electron; and
the state of the centre of mass of the total system.  The first two of
these are not necessarily separable from each other, but they are separable
from the third.  We can reasonably expect the state of the body plus
electron to have the following properties.
\renewcommand{\theenumi}{\roman{enumi}}
\begin{enumerate}
\item  The electron and the body are localised relative to each other.
\item  The total momentum of the body and electron in the centre of
mass frame is described by a momentum eigenstate of momentum $0$.
\item  The relative kinetic energy of the body and electron, plus the
internal energy of the body, is described by an energy eigenstate of energy
$E$.
\item  The centre of mass state of the total body plus electron system is
the same as the original centre of mass state of the body.
\end{enumerate}
Requirement~i is demanded by our normal empirical observation that particles
emitted in situations like this are localised.  Requirements~ii to iv are
demanded by the relevant conservation laws, together with the technique of
separation of internal and centre of mass states for systems.

	It may appear at first sight that the condition i is incompatible
with conditions ii and iii:  the Heisenberg uncertainty relations between
position and momentum appear to forbid i and ii from being realised at once.
The same intuition should also apply to the combination of i and iii, since
the state involves free particle kinetic energy eigenstates, which essentially
are the momentum eigenstates.  This appearance will turn out to be misleading,
as it must do.  If i, ii and iii cannot be combined then either energy-momentum
conservation or the known behaviour of emitted particles will be violated.

	We have the standard theorem that Hamiltonian time-evolution preserves
the mean value and distribution of the energy of a quantum system if the
Hamiltonian is time independent,\footnote{Messiah\cite{mess}, p195 and p210.
It is surprising how few texts contain this fundamental result.} which it is
presumed to be for an isolated system.  This theorem assures us that ii and
iii must hold, and if they fail then the status of conservation laws in
quantum mechanics will be called into question.

	The main argument of this paper proves that a quantum state conforming
to the above requirements can exist by constructing an example of such a
state.  The example is not the most physically natural such state, it is
chosen to make the mathematical and conceptual analysis as simple as
possible.  It may be helpful to briefly describe the course of the argument
in prose.

	I can give an intuitive feel for the form a state like this will
take.  To begin with I consider only the relative position and momentum
requirements (i and ii).  When the electron is emitted it has a momentum
distribution, which for the moment I will assume to be Gaussian.  Conservation
of momentum demands that the body recoil with equal and opposite momentum,
which amounts to a requirement that the body's momentum be anticorrelated
with that of the electron.  So the two momenta are anticorrelated and
Gaussian distributed.  The total momentum of the state is clearly zero
in the centre of mass frame, so the state is an eigenstate of momentum.
After transforming from the momentum basis to the position basis the
relative position of the electron and body will be Gaussian distributed.

	Now, this state is so far not an energy eigenstate:  anticorrelation
of the momenta still leaves a state of variable kinetic energy.  The residual
energy, the difference between the kinetic energy and the total energy $E$,
remains in the internal state of the body.  Hence the internal state cannot
be an energy eigenstate, it must involve a distribution of energies
anticorrelated with the kinetic energy of the body-electron state.  I will
be obliged to change the momentum distribution:  the internal energy
has a lower bound fixed by the lowest available internal energy state.
This will impose an upper bound on the kinetic energy.

	I will try to move toward the expected state in stages; first by
combining the localisation requirement i with the momentum requirement ii.
Then I will add the energy conservation requirement.  Since only the form of
the solutions is important, I set $\hbar = c = 1$ and set all normalisation
constants to one, without any loss of generality.

\section{Momentum and Localisation}

	In this section I find a state, $\ket{LK}$ (for localised
$k$-eigenstate), that satisfies only the localisation and momentum
conservation requirements above.  It is easy to write down the state
satisfying requirement ii; it can be written directly in the momentum
basis as a momentum anticorrelated state. I will assume a Gaussian
distribution of momentum, this being the normal choice for free particle
localised states.  It is then necessary to change to the position basis
to see whether the state has the desired localisation.

\begin{equation}
\ket{LK}=
\int_{k=-\infty}^\infty {
        e^{-k^{2}/{\sigma_k^2}} e^{i k x_{12}}
        \ket{k}_1 \otimes \ket{-k}_2
  dk }
\end{equation}

Where $\sigma_k$ is the standard deviation of the $k$-distribution and
$x_{12}$ is the mean separation of the bodies, 1 and 2, which determines
a relative phase factor between $k$-eigenvectors.  In general, integer
subscripts identify which body the symbol refers to.  To confirm formally
that this is an eigenstate of momentum, apply
\begin{equation}
  {\bf P}_{12} = {\bf P}_{1} \otimes I_2 + I_1 \otimes {\bf P}_{2}
\end{equation}
where ${\bf P}_{12}$ is the total momentum operator, ${\bf P}_{1}$ is the
momentum operator associated with body one, $I_2$ is the identity operator
associated with body two, and so on.

	Changing to the position basis gives

\begin{equation}
\ket{LK}=
\int_{x_1} {
  \int_{x_2} {
    \int_{k=-\infty}^\infty {
                e^{-k^2/\sigma_k^2}
                e^{i k \left( x_1-x_2+x_{12}\right)}
     dk }
     \ket{x_1}_1 \otimes \ket{x_2}_2
    dx_1 }
  dx_2 }
\end{equation}
(Where integration limits are not stated they are presumed to be from
$-\infty$ to $\infty$.)  This integral can be found from the standard
Fourier transform of a Gaussian, hence
\begin{equation}
= \frac{2\sqrt{\pi}}{\sigma_x}
\int_{x_1} {
  \int_{x_2} {
         e^{-\left(x_1-x_2+x_{12}\right)^2/\sigma_x^2}
          \ket{x_1}_1 \otimes \ket{x_2}_2
     dx_1 }
   dx_2 }
\end{equation}

This is a Gaussian localised state in $x_1-x_2$ with mean separation
$x_{12}$ and standard deviation $\sigma_x=2/\sigma_k$.  The two bodies
are localised relative to each other:  if (say) $x_1$ is fixed then $x_2$
will be known to within a Gaussian of mean $x_1+x_{12}$ and standard
deviation $\sigma_x$.  Localisation here is only relative; since $\ket{LE}$
is an eigenstate of total momentum it cannot be expected to contain any
centre of mass position information and it does not:  all values of $x_1$
are equally probable if $x_2$ is unknown.

	$\ket{LK}$ is a near descendant of the state described by Einstein,
Podolsky and Rosen;\cite{epr} it is in fact a real world version of
it.  EPR's state involved perfectly anticorrelated momenta and perfectly
correlated relative positions described by Dirac deltas.   This latter
was associated with an equal amplitude for all possible relative momenta.
EPR's state is an eigenstate of total momentum, meaning it contains no
centre of mass position information; and contains no relative momentum
information, allowing it to be a relative position eigenstate.  The new
state $\ket{LK}$ remains an eigenstate of total momentum and so leaves
the centre of mass position undefined, but the Gaussian distribution
imposed on the single particle momenta leaves the positions correlated
only up to a Gaussian.

	So $\ket{LK}$ is a state that satisfies the first two requirements
of momentum conservation and relative localisation.  I have chosen to
leave energy conservation aside for now, but  the final requirement,
that of preserving the centre of mass state, should apply.  Here there
is a problem because I have just argued that $\ket{LK}$ cannot contain
centre of mass position information:  $\ket{LK}$ is consistent only with
a completely delocalised centre of mass.  It is clear that if the centre
of mass position is localised the perfect anticorrelation of momentum
will be lost, so the state will no longer be a momentum eigenstate.

	Briefly, to prove this, consider a state, $\ket{LKL_{CM}}$,
obtained by imposing an effective centre of mass localisation on $\ket{LK}$
(and setting $x_{12}=0$ and $\sigma_x=1$ for simplicity)
\begin{eqnarray}
\ket{LKL_{CM}} &=&
\int_{x_1} {
  \int_{x_2} {
         e^{-x_1^2} e^{-\left(x_1-x_2\right)^2}
         \ket{x_1}_1 \otimes \ket{x_2}_2
     dx_1 }
   dx_2 } \nonumber \\
&=& \pi
\int_{k_1} {
  \int_{k_2} {
     e^{-k_2^2/4} e^{-(k_1+k_2)^2/4}
     \ket{k_1}_1 \otimes \ket{k_2}_2
  dk_1 }
dk_2 }
\end{eqnarray}
(by repeated use of integrals 3.923 in Gradshteyn\cite{grad}, p.520)
This state is clearly not anticorrelated in momentum and is therefore
not a momentum eigenstate.

	States like $\ket{LK}$ may certainly exist in our universe, but
they cannot describe the ordinary objects we see around us, because
ordinary objects have localised centres of mass.  A pair of bodies in
the state $\ket{LK}$ would be unobservable to us, and it seems unlikely
that bodies could remain in such a state for long.  Interaction with the
background electromagnetic radiation would eventually cause them to
become localised relative to us, a process that should occur under any
of the current competing theories of measurement.  Hence $\ket{LK}$ does
not solve the original problem of momentum conservation in particle
emission, and cannot be used to describe the centre of mass states of
normal objects.  It looks as though i and iv are not consistent with
each other.

	There is an alternative view of the situation, however.  If
the universe held only two planets then $\ket{LK}$ would be a good
description for them consistent with our expectations.  In particular,
the lack of centre of mass localisation is just what we expect if we
believe that position is meaningful only relative to other objects.
To a person standing on one of the two planets, the other would always
be localised while the centre of mass position would be unobservable.

	$\ket{LK}$ is not a useful description of bodies in our universe,
but it is an acceptable description of a universe containing only two
bodies.  Unfortunately, therefore, it does not address the problem
originally posed.  The question immediately occurs of whether $\ket{LK}$
can be extended to describe universes with larger numbers of bodies,
with all bodies pairwise relatively localised.  In this case there would
be the original two bodies plus a third where an external observer could
stand.  The first two bodies would be each localised relative to the
external observer as well as to each other, so the centre of mass of these
two would be effectively localised.  The centre of mass of the total
three-body system would not be localised, but again this is no problem
as there is no external object to which it could be related.

	I have reached the point of proposing a single extended form of
$\ket{LK}$, which I will call $\ket{LKn}$, encompassing every body in an
$n$-body universe.  It would have the two remarkable properties of being
a momentum eigenstate and of having no absolute centre of mass
localisation.  To prove that such a state is possible, begin with
the three body state, $\ket{LK3}$.

\begin{equation}
\ket{LK3} =
\int_{k_1} {
    \int_{k_2} {
        e^{-k_{1}^2/\sigma_k^2} e^{i k_1 x_{13}}
        e^{-k_{2}^2/\sigma_k^2} e^{i k_2 x_{23}}
        \ket{k_1}_1 \otimes \ket{k_2}_2
                  \otimes \ket{-k_1-k_2}_3
     dk_1 }
 dk_2 }
\end{equation}

This is a momentum eigenstate of eigenvalue $0$.  To check its
localisation properties transform to the position basis.

\begin{displaymath}
=\int_{x_1} {
  \int_{x_2} {
    \int_{x_3} {
      \int_{k_1} {
        e^{-k_{1}^2/\sigma_k^2} e^{i k_1 \left( x_1-x_3+x_{13}\right)}
        \int_{k_2} {
            e^{-k_{2}^2/\sigma_k^2} e^{i k_2
                \left(x_2-x_3+x_{23}\right)}
	dk_2 }
      dk_1 } } } } \times
\end{displaymath}
\begin{equation}
 \times \ket{x_1}_1 \otimes \ket{x_2}_2 \otimes \ket{x_3}_3
    dx_1
  dx_2
dx_3
\end{equation}
Applying the standard Fourier transform twice in succession gives
\begin{eqnarray}
= \frac{4 \pi}{\sigma_x^2}
\int_{x_1} {
  \int_{x_2} {
    \int_{x_3} {
        e^{-\left(x_1-x_3+x_{13}\right)^2/\sigma_x^2}
        e^{-\left(x_2-x_3+x_{23}\right)^2/\sigma_x^2}
      \times} } }
 \nonumber \\
    \times \ket{x_1}_1 \otimes \ket{x_2}_2 \otimes \ket{x_3}_3
           dx_1 dx_2 dx_3
\end{eqnarray}

 $\ket{LK3}$ is a state Gaussian localised in $x_1-x_3$ and in
$x_2-x_3$.  If any one of the variables $x$ is known then the other two
will be Gaussian localised.  If any one body is traced out, the
remaining two will be relatively Gaussian localised.  $\ket{LK3}$
therefore has similar properties to $\ket{LK}$; it is a momentum
eigenstate, its three bodies are relatively localised and its centre
of mass state has no localisation.

	With $\ket{LK}$ and $\ket{LK3}$ I can now give a description
of the thought experiment of section 1.  The experiment begins with
a two body universe described by $\ket{LK}$, one body being the original
object ready to emit an electron, the other a notional external
observer.  Once the electron has been emitted the description shifts
to $\ket{LK3}$ with object, electron and external observer.  The object
and the electron are each localised relative to the external observer,
so their joint centre of mass must be also.  There is no longer
anything to prevent this final centre of mass state from being the
same as the initial position state of the object alone, so requirement
iv can be met.  I know of no reason why the `external observer'
should not be the centre of mass of the rest of our universe, in
which case the experiment has been brought into the real world.
The following is equivalent to a proof that this is indeed possible.

	It is clear that the form of $\ket{LK3}$ can be extended to
any arbitrary number of particles $n$.

\begin{eqnarray}
\lefteqn{
\ket{LKn} =
\int_{k_1} { \ldots
    \int_{k_{n-1}} {
        e^{-k_{1}^2/\sigma_k^2} e^{i k_1 x_{1n}} \ldots
        e^{-k_{n-1}^2/\sigma_k^2} e^{i k_2 x_{n-1,n}}
     }
 } \times
} \nonumber \\
& & \times \ket{k_1}_1 \otimes \ldots \otimes \ket{k_{n-1}}_{n-1}
                  \otimes \ket{-k_1- \ldots -k_{n-1}}_n
    dk_1  \ldots
    dk_{n-1}
\end{eqnarray}

\begin{eqnarray}
= \left(\frac{2\sqrt{\pi}}{\sigma_x} \right)^{n-1}
\int_{x_1} { \ldots
  \int_{x_n} {
        e^{-\left(x_1-x_n+x_{1n}\right)^2/\sigma_x^2} \ldots
        e^{-\left(x_{n-1}-x_n+x_{n-1,n}\right)^2/\sigma_x^2}
  }
 } \times \nonumber \\
   \times \ket{x_1}_1 \otimes \ldots \otimes \ket{x_n}_n
  dx_1  \ldots
dx_n
\end{eqnarray}
A state like this can encompass the whole universe, therefore a
universe containing many localised particles can nevertheless be in
a momentum eigenstate.  Within this universe it seems that the thought
experiment, and processes like it, can occur meeting requirements i,
ii and iv.

	The next step is to make the state an energy eigenstate as
well.  The resulting state, $\ket{LKE}$, contains a third component
representing the internal state of one or more of the bodies.  The
energy of the internal state is anticorrelated with the kinetic energy
of the free bodies, so the total energy of the whole state is a
well-defined constant $E$.   Before doing this I want to bring out some
mathematical and physical ramifications of that step by trying an
analogous thing in a much simpler setting.

\section{An Analogy in Spin States}
	Consider the correlated spin state written in the $z$-component
basis
\begin{equation}
\ket{XZ} = \uk_1 \otimes \dk_2 + \dk_1 \otimes \uk_2
\end{equation}
Changing to the $x$-component basis this is
\begin{equation}
\ket{XZ} = \rk_1 \otimes \rk_2 - \lk_1 \otimes \lk_2
\end{equation}
That is, $\ket{XZ}$ is anticorrelated in the $z$-component basis and
correlated in the $x$-component basis, analogously to $\ket{LK}$.  I
want to explore the problem of creating $\ket{LKE}$, which is $\ket{LK}$
but with the addition of a third component whose energy is anticorrelated
with the kinetic energy of the free bodies, leaving the total energy
constant.  Analogously, I will attempt to put a third spin in correlation
with the possible states in the $z$-component basis, as in
\begin{equation}
  \ket{XZE} = \uk_1 \otimes \dk_2 \otimes \dk_3 +
                \dk_1 \otimes \uk_2\ \otimes \uk_3
\end{equation}
If this state works the way $\ket{LKE}$ is expected to, a change to the
$x$-component basis will produce $\ket{XZE}$ something like
\begin{equation}
   \rk_1 \otimes \rk_2 \otimes \ket{s}_3 -
   \lk_1 \otimes \lk_2 \otimes \ket{t}_3
\label{3.4}
\end{equation}
where $\ket{s}_3$ and $\ket{t}_3$ are two states in $\lk_3$ and $\rk_3$.

	However, the result of changing basis is no such thing.
\begin{eqnarray}
  \lefteqn{\ket{XZE}} \nonumber \\
 &=& \frac{1}{2\sqrt{2}}(\rk_1 + \lk_1) \otimes (\rk_2 - \lk_2) \otimes
        (\rk_3 - \lk_3) + \nonumber \\
 & &  \;(\rk_1 - \lk_1) \otimes (\rk_2 + \lk_2) \otimes (\rk_3 + \lk_3)
     \nonumber \\
 &=& \frac{1}{\sqrt{2}}((\rk_1 \otimes \rk_2 \otimes \rk_3) - (\lk_1 \otimes
       \lk_2 \otimes \rk_3) + \nonumber \\
 & & \;\;\;\;(\rk_1 \otimes \lk_2 \otimes \lk_3) - (\lk_1 \otimes \rk_2 \otimes
      \lk_3) )
\label{3.5}
\end{eqnarray}
Which is not the required form.  In particular, 1 and 2 are uncorrelated.

	Mathematically, this comes about because the final form of
$\ket{XZ}$ in the $x$-component basis is produced by cancellation
of $x$ terms arising from different terms in the $z$-component
representation.  In $\ket{XZE}$ the $z$ terms of $\ket{XZ}$ are
multiplied by new factors that differ, so it is not to be expected
that the required cancellation will occur.  Exploration with alternative
forms for the ``energy'' states suggests that there is no choice that
produces the required result.  This does not augur well for the next
calculation:  $\ket{LK}$ surely has the equivalent property, that terms
of different energy contribute to cancellations when changing basis.
Nevertheless, there are many important disanalogies between the spin and
position-momentum cases, and this particular result does not carry over.

	There is an important point that will carry over, however.
Suppose for a moment that (\ref{3.4}) were the correct form for
$\ket{XZE}$ in the $x$-component basis.  Now consider what happens
if the $z$-component of spin~3 is measured.  Whichever result appears
the anticorrelated state is broken up and the positive correlation in
the $x$-component basis is destroyed.  For instance, assume the result
is $\dk_3$.  Then the new state $\ket{M}$ will be
\begin{eqnarray}
\ket{M} &=& \uk_1 \otimes \dk_2 \otimes \dk_3 \nonumber \\
  &=& \frac{1}{2\sqrt{2}}(\rk_1 + \lk_1) \otimes (\rk_2 - \lk_2)
                            \otimes ((\rk_3 - \lk_3) \nonumber \\
  &=& \frac{1}{2\sqrt{2}}(\rk_1 \otimes \rk_2 \otimes \rk_3) -
       (\rk_1 \otimes \rk_2 \otimes\lk_3)- \nonumber \\
  & & (\rk_1 \otimes \lk_2 \otimes \rk_3) +
       (\rk_1 \otimes \lk_2 \otimes\lk_3) +
       \nonumber \\
  & & (\lk_1 \otimes \rk_2 \otimes \rk_3) -
       (\lk_1 \otimes \rk_2 \otimes\lk_3) -
       \nonumber \\
  & & (\lk_1 \otimes \lk_2 \otimes \rk_3) +
       (\lk_1 \otimes \lk_2 \otimes\lk_3)
\end{eqnarray}
Which is easily distinguished from (\ref{3.4}) by measurements made on
spins $1$ and $2$ alone.  In (\ref{3.4}) only anticorrelated results are
possible, while in $\ket{M}$ the other combinations are equally likely.
Hence, if $\ket{XZE}$ really had the form (\ref{3.4}) in the $x$-component
basis it would provide the core of a superluminal communicator.  An
experimenter with access to $1$ and $2$ could tell whether a second
experimenter at a different place had measured $3$.  As I have shown,
$\ket{XZE}$ cannot be used as a communicator because its correct form
is (\ref{3.5}) rather than (\ref{3.4}).  In the next two sections it
will become clear that $\ket{LKE}$ does have a form which allows it
in principle to be used as a superluminal communicator.

\section{Energy, Momentum and Localisation}

	In this section I will construct the state $\lke$, which is an
extension of $\LK$ to an energy eigenstate.  I will construct an
energy-momentum eigenstate in the momentum basis, and then show that
it is a position localised state by transforming to the position basis.
In this case, the integrals involved are much more difficult, and it
will be necessary to adopt a set of simplifying approximations, both
in the form of the states carrying the residual energy and the
distribution over $k$.  These approximations should not have any
substantial effect on the physics of the final result, but even with
their help, it will be necessary to evaluate the integrals by numerical
means.

	The first question to be addressed is the form of the states
that will carry the residual energy.  In the original electron emission
example the residual energy would remain in the internal states of the
electron gun itself, somewhere in the conduction or valence bands of
the metal.  I am looking for a proof in principle only, so I want to
avoid the mathematical difficulty of a full model for the internal
states.  I will adopt the simplest possible approximation having the
necessary properties.  These are:
\renewcommand{\theenumi}{\alph{enumi}}
\begin{enumerate}
   \item continuous energy eigenvalues,
   \item zero momentum.
\end{enumerate}
The internal energy states of solids have both these properties.
Even though confined similarly to a particle in a box, the valence
and conduction electrons of a solid have available continuous bands
of energy levels.  And having zero total momentum is a defining property
for an internal state:  the total momentum of an isolated body appears
in its centre of mass state; the internal momentum must be zero.

	The simplest continuum energy state is an infinite plane standing
wave, but this is not a zero momentum state, although its mean momentum
is zero.  To get zero momentum we will need to consider a complete
two-particle bound state.  The relationship between the two particles
of a bound state is very similar to that of the two free particles in
$\LK$ itself:  at any instant their momenta are equal and opposite.
I will make the drastic simplifications of not considering the force
binding the two particles together, and of not confining them to a
finite volume.  The model this leaves is of two particles in a
momentum-anticorrelated infinite standing wave.  The energy state
will be
\begin{eqnarray}
\ket{E^\prime} & = &
\ket{k^\prime}_1 \otimes \ket{-k^\prime}_2 +
    \ket{-k^\prime}_1 \otimes \ket{k^\prime}_2 \nonumber \\
& = &
\int_{y_1} {
  \int_{y_2} {
    \left( e^{i k^\prime (y_1 - y_2)} + e^{-i k^\prime (y_1 - y_2)} \right)
    \ket{y_1}_1 \otimes \ket{y_2}_2
  dy_1 }
dy_2 } \nonumber \\
& = &
\int_{y_1} {
  \int_{y_2} {
    \cos k^\prime (y_1 - y_2)
    \ket{y_1}_1 \otimes \ket{y_2}_2
  dy_1 }
dy_2 }
\end{eqnarray}
Where $y_1$ and $y_2$ are the position variables for the two particles
and $k^\prime = \sqrt{E^\prime}$ (With mass set to $1$).  (Recall that
$\hbar = c = 1$.)  The residual energy for a given value of $k$ in
$\LK$ is $E - k^2$, where $E$ is the total energy (and the masses
of the free bodies have also been set to $1$).  Hence, the internal
energy state for a given value of $k$ is
\begin{equation}
\ket{E-k^2} =
\int_{y_1} {
  \int_{y_2} {
    \cos \left( (y_1 - y_2) \sqrt{E-k^2} \right)
    \ket{y_1}_1 \otimes \ket{y_2}_2
  dy_1 }
dy_2 }
\end{equation}

	The next question that must be decided is the distribution
over $k$ for $\lke$.  This is a problem because energy can only be
positive.  A large value of kinetic energy in the free bodies cannot
be balanced by a negative value of the residual energy.  There must
be an upper bound to the energy, and hence an upper bound to the
absolute value of $k$.  The combination of the energy factor and
the finite range of $k$ means that the change-of-basis integral for
$\lke$ will be much more difficult to solve than that of $\LK$.
To keep the problem as simple as possible I will choose a square
distribution for $k$.  The final form of $\lke$ is therefore

\begin{eqnarray}
 \ket{LKE} =
\int_{k=-k_0}^{k_0} {e^{i k x_{12}}
        \ket{k}_1 \otimes \ket{-k}_2 \otimes \left(
           \ket{\sqrt{E-k^2}}_3 \otimes \ket{-\sqrt{E-k^2}}_4 + \right.
        } \nonumber \\
     \left. +\ket{-\sqrt{E-k^2}}_3 \otimes \ket{\sqrt{E-k^2}}_4
     \right)
    dk
\label{4.3}
\end{eqnarray}
where $E \ge k_0^2$ is the total energy of the state.  Now transform
to the position basis.
\begin{eqnarray}
\ket{LKE} =
\int_{x_1} {
  \int_{x_2} {
    \int_{y_3} {
      \int_{y_4} {
        \int_{k=-k_0}^{k_0} {
            e^{i k \left( x_1-x_2+x_{12}\right)}
            \cos( \sqrt{E-k^2} \left( y_3-y_4\right))
         dk \times }}}}}
       \nonumber \\
       \times \ket{x_1}_1 \otimes \ket{x_2}_2
              \otimes \ket{y_3}_3 \otimes \ket{y_4}_4
      dx_1
    dx_2
  dy_3
dy_4
\end{eqnarray}
with $E \ge k_0^2$.

	The integral in $k$ can be simplified in two ways.  First, by
substituting $x=x_1-x_2+x_{12}$ and $y=y_3-y_4$.  Second, by
substituting $e^{i t}=\cos t + i \sin t$ and noting that the imaginary
part is antisymmetric and therefore its integral, over symmetric
limits, is zero.  This leaves
\begin{equation}
  \Psi(x,y)=
    \int_{k=-k_0}^{k_0} { \cos (xk) \cos \left(y \sqrt{E-k^2}\right) dk }
\label{4.5}
\end{equation}
as the integral that must be solved.  This form has two close relatives
in Gradshteyn (\cite{grad} 3.711 p439, 3.876-7 p508), but, despite
considerable effort on the problem, I was forced to resort to
numerical analysis to find the form of its solution.  The calculation
was performed using the quadrature routines of the Nag Fortran
Library version 16; specifically, routine D01AKF, running on a Sun
SPARCstation IPC.

	Figure~1 shows the result of integral~(\ref{4.5}) evaluated
for $k_0=1$, $E=1$, over a range of $-25$ to $25$ for both $x$ and
$y$, at a spacing of $1$.  This plot was generated using Mathematica~2.2,
as were those following.  It is clear that the state is localised in $x$,
though not as sharply as it would be if the $y$ subsystem did not
exist.  In fact, for $y=0$ the integral is exactly what it would be
if the residual energy states did not exist.  However, for values of
$x$ away from zero the maximum amplitude is not at $y=0$ but at a value
near to $y=|x|$.  This means that values of $x$ away from zero are more
probable overall than they would be without the residual energy
component, and so the localisation is less sharp.  Nevertheless,
barring unexpected behaviour outside the region so far explored,
which is very unlikely given the good behaviour of the two related
integrals above, there is little doubt that the system is localised
in $x$.

	The really important issue here is the relative localisation
of $x_1$ and $x_2$, irrespective of the behaviour of $y_3$ and $y_4$.
To display the $x$-localisation more clearly I trace out the internal
states, which amounts to squaring the result of the first integration
and then integrating over $y$ for each value of $x$.  This has the
effect of averaging over $y$.  Values were calculated over the
range $-40$ to $40$, at a spacing of $.25$, for $x$ and $y$, then
the integration over $y$ was performed using Nag routine D01GAF.
The resulting unnormalised position distribution $d(x)$ is displayed
in figure~2.  The system is localised in $x$, and hence is relatively
localised in $x_1$, $x_2$ with mean separation $x_{12}$.

	The state $\lke$ fulfils the requirements i - iv set out in
section one.  Equation (\ref{4.3}) clearly shows it is an energy-momentum
eigenstate, while figure~2 shows that it is well localised.
Remembering that our expectations of requirement~iv have been altered
by interpreting $\lke$ as describing an entire two-body universe,
the state has all the properties demanded of it.  $\lke$ is a
position-localised, energy-momentum eigenstate.

	The numerical nature of the above calculation makes it impossible
to generalise directly to arbitrary numbers of particles, as needs to
be done to sustain the interpretation of $\lke$ as encompassing the
entire universe.  The fact that a state like $\lke$ exists at all is
telling, and given also that $\LK$ generalises to $\ket{LKn}$, there
is every reason to expect that $\lke$ will do likewise.

	As a check, I will do the calculation for the next case,
$\ket{LKE3}$.  The state is

\begin{eqnarray}
\ket{LKE3} =
\int_{k_1=-k_0}^{k_0} {
  \int_{k_2=-k_0}^{k_0} { e^{i k_1 x_{13}} e^{i k_2 x_{23}}
        \ket{k_1}_1 \otimes \ket{k_2}_2 \otimes
        \ket{-k_1-k_2}_3 \otimes }} \nonumber \\
	\otimes \left( \ket{\sqrt{E-E_k}}_4
	\otimes \ket{-\sqrt{E-E_k}}_5 + \right. \nonumber \\
    +	\left. \ket{-\sqrt{E-E_k}}_4
	\otimes \ket{\sqrt{E-E_k}}_5 \right)
  dk_1
dk_2
\end{eqnarray}
with $E \geq 3k_0^2$ and $E_k=(1/2)(k_1^2+k_2^2+(k_1+k_2)^2)$ (Again
with all $m=1$).
Change of basis:
\begin{eqnarray}
= \int_{x_1} {
  \int_{x_2} {
    \int_{x_3} {
      \int_{y_4} {
        \int_{y_5} {
          \int_{k_1=-k_0}^{k_0} {
          \int_{k_2=-k_0}^{k_0} {
            e^{i \left( k_1 \left( x_1-x_3+x_{13}\right) +
             k_2 \left( x_2-x_3+x_{23}\right)\right)} }}}}}}}
                                          \times \nonumber \\
  \times
     \cos\left(\left(y_4-y_5\right)
             \sqrt{E-\frac{1}{2}({k_1}^2+{k_2}^2+(k_1+k_2)^2)}\right)
     dk_1
     dk_2 \times \nonumber \\
  \times  \ket{x_1}_1 \otimes \ket{x_2}_2 \otimes \ket{x_3}_3
       \otimes \ket{y_4}_4
       \otimes \ket{y_5}_5
      dx_1
    dx_2
    dx_3
  dy_4
dy_5   \label{4.7}
\end{eqnarray}
with $E \geq 3k_0^2$.
To solve the integral in $k_1$, $k_2$ make equivalent simplifications to
those for $\lke$.  Substitute $xx_1=x_1-x_3+x_{13}$, $xx_2=x_2-x_3+x_{23}$
and $y=y_1-y_2$; and substitute $e^{i t}=\cos t + i \sin t$.  Again,
the imaginary part is zero because of its integrand's symmetry
properties.  This can be seen by dividing the integration range
into quadrants:
\begin{eqnarray}
  \int_{k_1=-k_0}^{k_0} {
  \int_{k_2=-k_0}^{k_0} { \ldots }} =
     \int_{k_1=0}^{k_0} {
     \int_{k_2=0}^{k_0} { \ldots }} +
     \int_{k_1=-k_0}^{0} {
     \int_{k_2=-k_0}^{0} { \ldots }} + \nonumber \\
   + \int_{k_1=0}^{k_0} {
     \int_{k_2=-k_0}^{0} { \ldots }} +
     \int_{k_1=-k_0}^{0} {
     \int_{k_2=0}^{k_0} { \ldots }}
\end{eqnarray}
The first two of these terms cancel, as do the third and fourth.

	This integral was solved using Nag routine D01DAF for two
dimensional integrals over a product region, using the values $k_0=1$,
$E=3$.  Solutions were generated for $0 \le xx_1 \le 40$, $-20 \le xx_2
\le 20$ at intervals of $1$.  This is sufficient since the integral is
clearly symmetric on reflection in the lines $xx_1=|xx_2|$.  The
range of $y$ was $-40 \le y \le 40$ at intervals of $.25$.  The
internal energy states were traced out by integrating over $y$ the
square of the result for each value of $xx_1$ and $xx_2$, using
routine D01GAF as before.  The resulting unnormalised position
distribution $d(xx_1,xx_2)$ is displayed in figure~3.

	Figure 3 shows a state localised in $xx_1=x_1-x_3+x_{13}$ and
$xx_2=x_2-x_3+x_{23}$, and hence having relative localisation in
$x_1$, $x_2$ and $x_3$.  The three particle state has the required
properties, it is an energy-momentum eigenstate with all three
particles localised relative to each other.

	I consider the expectation that $\lke$ will generalise to
$\ket{LKEn}$ for arbitrary values of $n$ to be a good one.  For the
purposes of the remaining argument I will assume that $\ket{LKEn}$
exists and has equivalent properties to $\lke$ and $\ket{LKE3}$,
while admitting that a proof of this has not been given.

\section{Nonlocality}
	Section three was an investigation of the possibilities of
adding a third correlate into a two-particle correlated state.  As
such it was not very informative.  But the discussion there raised
the possibility that three-component correlated states might be
nonlocal, and nonlocal in a way allowing transmission of information.
Since it turns out that $\lke$ is a three-component correlated state
analogous to the one envisioned in section three, I will look in it
for the expected nonlocality.

	The nonlocality is easy to demonstrate.  As figure~2 shows,
$\ket{LKE}$ is well localised in position.  It is clear from the
momentum representation of $\ket{LKE}$ that the residual energy
eigenstates are in two to one correspondence with the free particle
momentum eigenstates for bodies $1$ and $2$ (the two momentum states
are identical with $k_1$ and $k_2$ interchanged).  If the residual
energy were to be measured exactly the free bodies would be left in a
superposition of these two momentum eigenstates.  The momentum
eigenstates are completely delocalised, and this change would be
readily detectable to any observer.

	If the residual energy were all in the internal state of body
one, an observer on body one could signal an observer on body two by
measuring body one's internal energy.  This is a fairly drastic kind
of signal -- complete delocalisation of the universe -- but a signal
is a signal.  Even if the residual energy were divided between the
two bodies' internal states, a measurement of one internal energy
would narrow the distribution of the relative momentum, leading to
a broadening of the relative position distribution.  This effect is
in principle observable even if quite small.

	For $\ket{LKEn}$ the nonlocality is harder to see, but can be
shown to exist for any value of $n$, though its exact form is more
difficult to find.  The following analysis does not capture the full
extent of the nonlocality, but it is sufficient to prove the point.
Take $\ket{LKE3}$, equation~(\ref{4.7}), as an example.  If we measure
the residual energy and get the result $E_m$, the state will change
in several ways.  First, the residual energy factor is now independent
of $k_1$ and $k_2$ (it becomes  \(\cos((y_4-y_5)\sqrt{E_m}) \)),
and so drops out of the $k$-integral.  At the same time, $k_1$
and $k_2$ cease to be independent of one another; they are related by
\begin{equation}
  \frac{1}{2}\left(k_1^2 + k_2^2 + (k_1 + k_2)^2\right) = E - E_m
\label{6.1}
\end{equation}
with  $0 \le E_m \le E$.
The restriction is demanded by conservation of energy (recall that
$E$ has been chosen to equal the maximum possible kinetic energy
$3k_0^2$, if $E$ were larger the restriction would need to be
suitably modified).  The effect of this is to make the distribution
of the free body momenta narrower, necessarily broadening the
distribution of the positions of the three bodies.  In the extreme
case as $E_m$ approaches $E$ we have $k_1$, $k_2$, $k_3$ all
approach $0$:  the positions of all bodies will be completely
delocalised as the $k$-distribution becomes infinitely narrow.

	As before, measuring the residual energy produces an
observable effect elsewhere in the universe.  This effect is
chance dependent, since the amount of broadening depends on the
value found for $E_m$, but remains a real nonlocality despite
this.  Furthermore, it is clear that a similar effect will occur
for any number of free bodies:  when $E_m$ is close to $E$ all
the bodies will be substantially delocalised because of the
narrow distribution allowed for their kinetic energies.  Provided
$\ket{LKEn}$ has the expected properties for arbitrary $n$, a
detectable nonlocality can be manifested in a universe of arbitrary
size, clearly including one the size of ours.

	There remains one questionmark over this nonlocality.
The supposedly nonlocal behaviour occurs only after a measurement
has been performed on every object in the universe to find the total
residual energy.  It could be argued that this phenomenon is not a
true nonlocality:  that although the behaviour looks superficially
like nonlocality it is not capable of carrying signals.  That would
not be surprising; it is the most widely accepted interpretation of
the Bell and GHZ nonlocality theorems for spin states (see e.g.
d'Espagnat\cite{desp} and Greenberger et.al.\cite{ghz}).  The states
cannot be used to transmit signals even though they are nonlocal in
a formally defined sense.\footnote{Although it should be remembered
that a case for an underlying physical nonlocality could be made,
at least within a realist framework.\cite{sim1,sim2}}

	This argument is incorrect.  It depends on the assumption
that the residual energy is spread through the internal states of
all the bodies in the universe.  While this is most likely true for
our universe, it is in principle possible for all the residual energy
to be held within a confined volume, and then the delocalisation of
the bodies within the universe after its measurement would be
nonlocal beyond doubt.

	The important point of this section is the apparent existence
of a nonlocality allowing transmission of signals which is inherent
in states of the form $\lke$, and therefore inherent in quantum mechanics.

\section{Conclusion}
	This paper has described a new form of quantum state,
$\lke$, with two important properties.  First, it is a localised
energy-momentum eigenstate.  Second, it has the potential to exhibit
nonlocal behaviour in the positive sense of sending signals faster
than light.  These remarkable qualities are offset by an awkward
facet:  $\lke$ is not suitable in general for describing ordinary
isolated objects in our own universe because the centre of mass
position of objects so described cannot be localised.  The most
natural use for these states is to describe an entire universe,
encompassing every body contained in it.  Here the lack of localisation
of the centre of mass becomes a virtue, reflecting the non-absoluteness
of position.  The need to encompass an entire universe makes $\ket{LKEn}$
difficult to use, because calculations for it can so far only be done
numerically and no proof exists that $\ket{LKEn}$ has the correct
properties for all $n$, although a failure of this would seem unlikely.
The difficulty in analysing $\ket{LKEn}$ mathematically also explains
why I have only offered proofs in principle for the existence of these
states, rather than a more physically realistic treatment.

	Given the peculiar properties of states like $\lke$, why
should we imagine that our universe is in one?  I have no conclusive
arguments to offer for this, but I can offer some plausibility
arguments.  First of all, the idea must have a strong intuitive appeal.
Our expectations about energy-momentum conservation are most simply
and directly realised if the universe is in an energy-momentum
eigenstate.  In the case of momentum, it is not clear that non-zero
momentum for the centre of mass of the universe is a possibility we
can give meaning to.  It certainly is not consistent with any notion
of momentum as a relative quantity.  Only if the centre of mass state
of the universe is a momentum eigenstate of momentum zero is this
problem definitely dealt with.

	The equivalent argument does not hold for energy, since it
is not a relative quantity.  However, there is little doubt that the
total energy of the universe has consequences detectable from within;
certainly in the big bang model the total energy affects the history
and development of the universe in an observable way.  If the universe
is not an energy eigenstate, then its observable state may not be well
defined.

	A more subtle argument is that $\ket{LKEn}$ would fill a hole
in our understanding of energy conservation in quantum mechanics.  We
have the previously mentioned (see footnote 1) theorem that energy is
conserved, in the sense that its mean value and distribution are
conserved, by Hamiltonian time evolution under a time-independent
Hamiltonian operator.  Yet we are in a peculiar position in that many
of the detail processes we know of, such as photon emission (e.g.
Davies\cite{dave} p108), have not been shown to conserve energy
explicitly.  The universe model suggested by $\ket{LKEn}$ is one
in which we know why individual processes conserve energy, and
therefore why the universe as a whole does.  If the big bang theory
of cosmology is correct, then the intimate entanglement of all the
bodies in the universe is explained by their component particles being
able to trace, directly or indirectly, a history back to an epoch when
all the contents of the universe were in close interaction.

	None of these arguments is conclusive, but one important point
remains even if they are discounted.  In the past the possibility of a
universal energy-momentum eigenstate was not considered because there
was no realisation that such a thing was possible.  Now it seems the
possibility exists, and it deserves investigation.

	Leading naturally to the most important question, that of
experimental tests.  The first potential test of whether the universe
is in a state like $\ket{LKEn}$ is that localisation of the bodies is
broader than it would be for an isolated body having the same momentum
distribution.  The isolated body's wave function is the same as the
$y=0$ slice on the plot, which clearly has smaller amplitude for large
$x$ than that for $y \approx x$.  It might be possible to perform an
experiment to detect this broadening of position uncertainty.
Remember, however, that the $k$-distribution of $\lke$ was chosen
only for convenience of calculation.  It is certainly not the best
choice, and there is no telling how other distributions might affect the
discrepancy between the isolated body and $\lke$ position uncertainties.
Tests might also arise from the nonlocal properties of $\lke$, or from
cosmological consequences arising from the state.  I cannot speculate
about the form such tests might take.

	We have been handed a surprise in $\lke$ for a reason I have
not so far discussed.  When explaining the counterintuitive properties
of quantum mechanics one of the favourite examples has been the
electron two-slit interference experiment (Feynman,\cite{feyn} p1-5).
A point that is always emphasised is the fact that the interference
pattern vanishes if we attempt to determine which slit each electron
passed through.  This principle is generalisable to the idea that
whenever interference occurs between multiple channels, or pathways,
the interference exists only so long as it remains impossible to
determine which channel the system passed through.  The moment a
correlation is established that would allow us to determine which
channel the system occupied the interference effect vanishes.  This
behaviour shows up in many places, such as two particle double-slit
experiments.\cite{ghz2}  I have not been able to find a general proof
of this principle; it seems to be widely accepted on the lack of
counterexamples.

	The $\lke$ states form a counterexample.  The localisation
peak is an interference maximum formed by a sum over many momentum
states, each of which forms a separate channel.  Yet the momentum states
are correlated with the residual energy states, so that measuring the
residual energy determines which interference channel the free particle
occupies.  Apparently, it is possible to arrange a state with
interference channels in correlation with the measurable state of
some other system in certain circumstances.  Indeed, it is this very
fact that opens the possibility of superluminal signalling.

	The nonlocal character of this class of states has a more
immediate impact on quantum theory than its other properties.  Whether
the universe is in such a state or not, the existence of $\lke$ is a
proof in principle that quantum mechanics is nonlocal, and demands
either a general proof that such states cannot exist in our universe,
or a reappraisal of the relationship between quantum and relativity
theory.

\section*{Acknowledgements}

	I would like to thank T.J. Penttila and A. Rawlinson for very
valuable advice and discussions.

\clearpage
\psfig{file=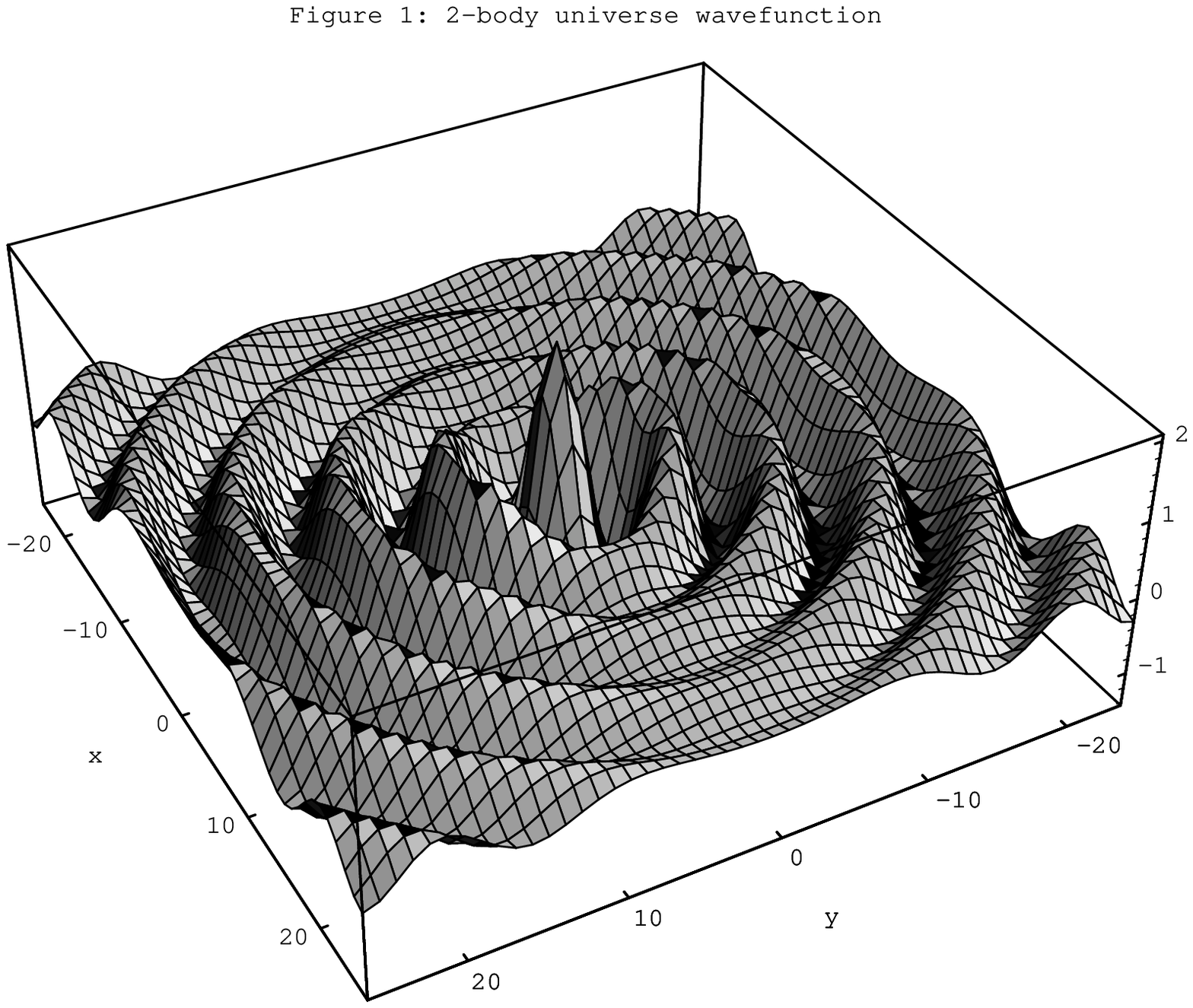,width=11.7cm}

\clearpage
\psfig{file=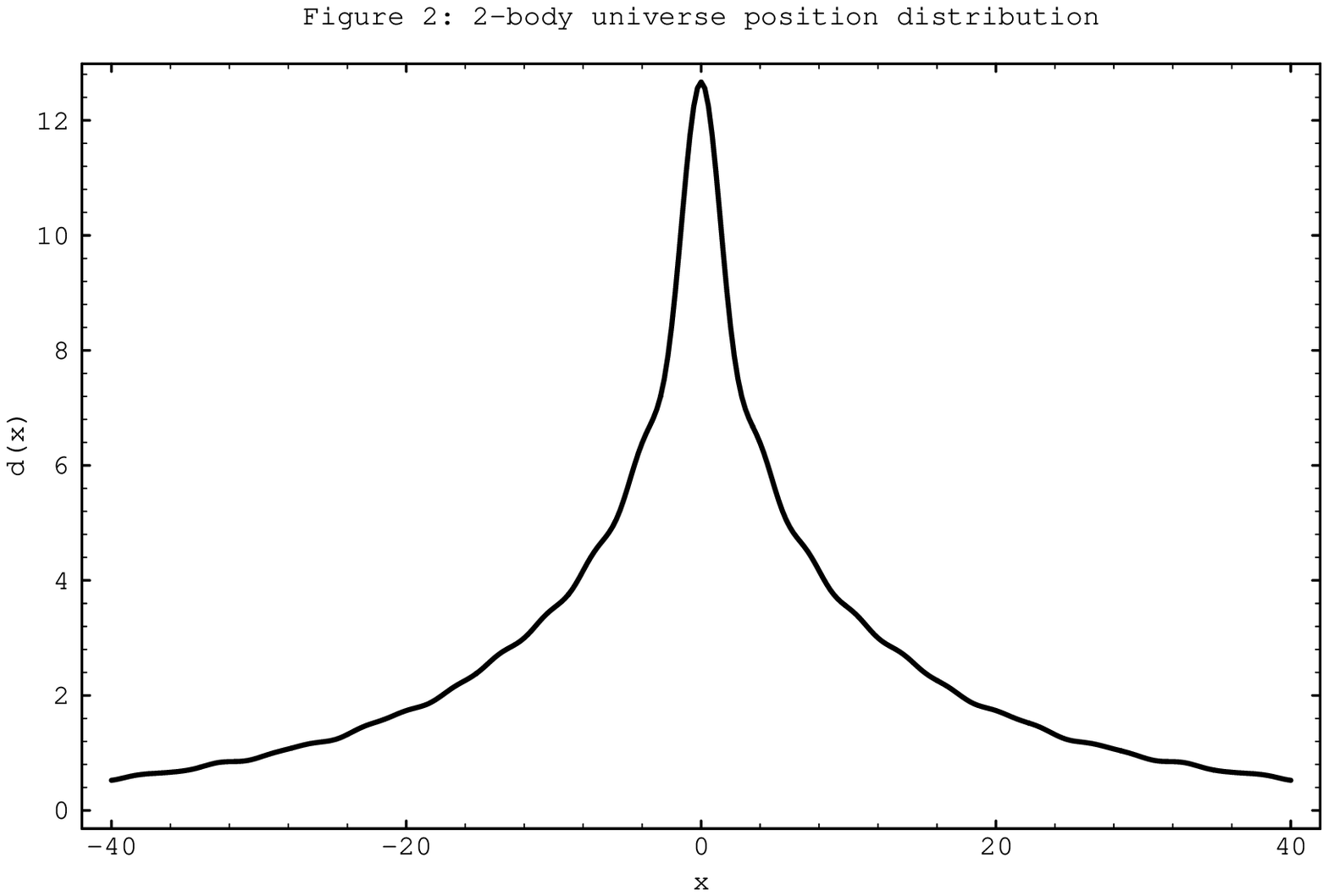,width=11.7cm}

\clearpage
\psfig{file=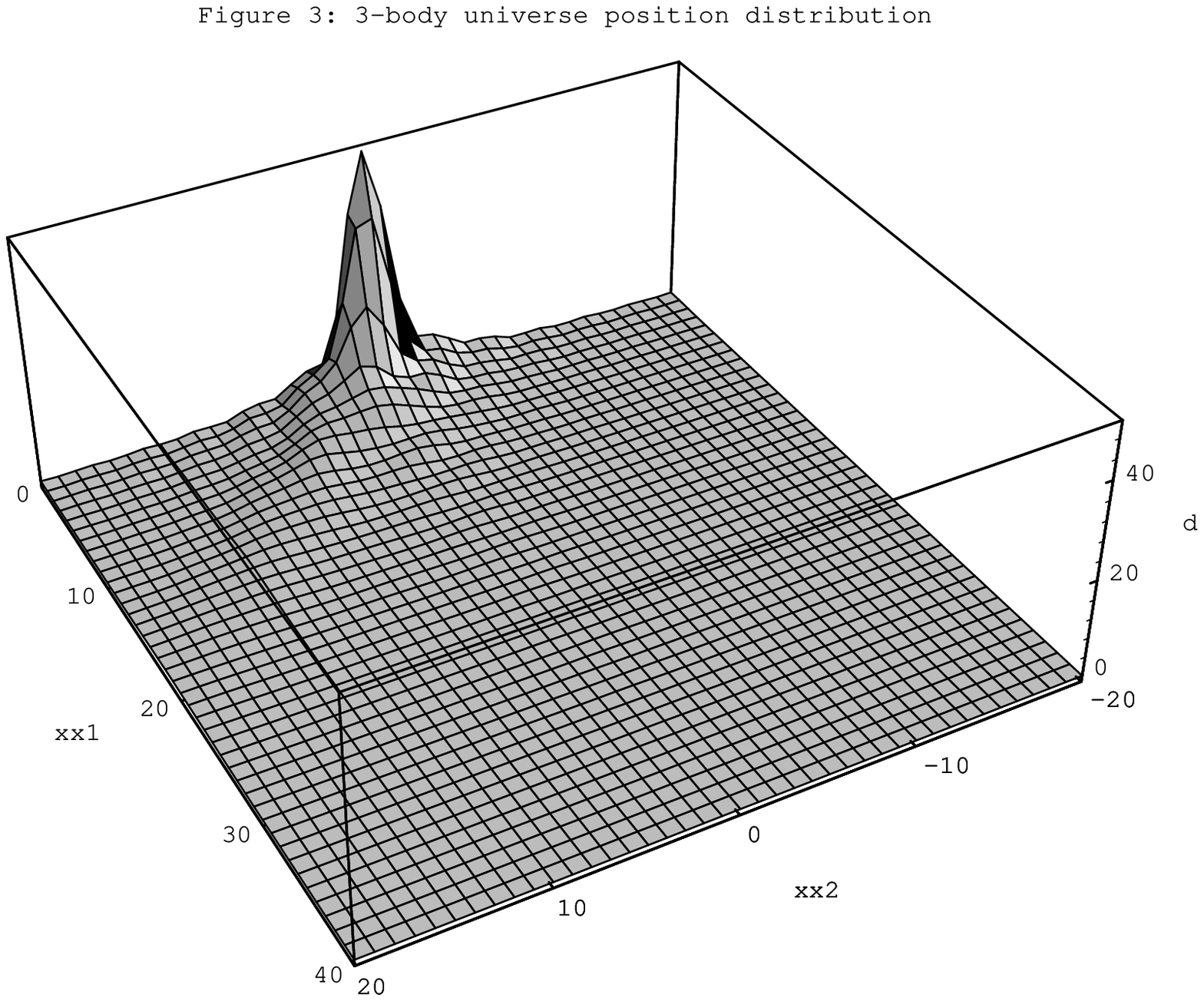,width=11.7cm}

\end{document}